\documentclass[aps, pra, reprint, floatfix, showpacs, footinbib]{revtex4-1}
 \usepackage[T1]{fontenc}
 \usepackage{graphicx}
 \usepackage{amsmath,amssymb}
\begin{document}
\title{Magnetically tunable Feshbach resonances in Li+Er}

\author{Maykel L.~Gonz\'alez-Mart\'{\i}nez}
\email{maykel.gonzalez-martinez@u-psud.fr}
\affiliation{Laboratoire Aim\'e Cotton, CNRS,
             Universit\'e Paris-Sud, ENS Cachan,\\
             B\^{a}t.\ 505, Campus d'Orsay,
             91405 Orsay, France}

\author{Piotr S.~\.{Z}uchowski}
\email{pzuch@fizyka.umk.pl}
\affiliation{Institute of Physics, Faculty of Physics, Astronomy and Informatics,
             Nicolaus Copernicus University,
             ul.\ Grudziadzka 5/7,
             87-100 Torun, Poland}

\date{\today}
\begin{abstract}
We explore the magnetic Feshbach spectra of ultracold ground-state Li+Er systems.
Our calculations predict many tunable resonances at fields below $1\,000$~G that
could be stably tuned in ultracold experiments.  We show that Li+Er spectra are
much less congested than those of systems involving heavier highly-magnetic atoms
and exhibit non-chaotic properties.
These features would facilitate identifying and addressing individual resonances.
We derive a simple model for the mass-scaling shifting of low-field resonances
that may simplify designing experiments with different Er bosonic isotopes.
Our work establishes Li+Er as very promising
systems for quantum simulation, precision measurements and the formation of polar 
paramagnetic molecules.
\end{abstract}


\maketitle
Ultracold species make it possible to build state-se\-lected quantum systems
with controllable interactions, which open the door to exploring fascinating
phenomena.  Among their many applications \cite{LDCarr:09b,ODulieu:09a},
ultracold systems can be used as quantum simulators
\cite{IBloch:12a,MABaranov:12}, to study condensed-matter physics
\cite{AMicheli:06,AMicheli:07,HPBuchler:07} and quantum-controlled chemistry
\cite{K-KNi:10,SOspelkaus:10b,MHGdeMiranda:11a}, to develop quantum information
devices \cite{ADerevianko:04,PRable:06} and ultraprecise spectroscopy
\cite{JJHudson:02,TZelevinsky:08,BJBloom:14}.

Tunable Feshbach resonances \cite{CChin:10} are powerful tools to control the
interaction and scattering properties of ultracold species, making many of these
applications possible.  Moreover, they are essential in the most successful
scheme to date to produce ultracold molecules: The magneto- or photoassociation
of ultracold atoms \cite{JMHutson:06,TKohler:06,KMJones:06} followed by coherent
\cite{FLang:08,K-KNi:08,JGDanzl:10,TTakekoshi:14a} transfer of the created 
molecules to their rovibrational ground states.  Having the possibility to address 
and tune across selected Feshbach resonances is thus key in ultracold experiments.

Recent advances in cooling highly-magnetic atoms such as Cr($^7$S)
\cite{AGriesmaier:05,ADePaz:13}, Dy($^5$I) \cite{MLu:11,MLu:12} and Er($^3$H)
\cite{KAikawa:12,KAikawa:14a} open exciting opportunities for tunability and
control \cite{SKotochigova:14a}.  The interaction between these atoms, however,
leads to highly congested Feshbach spectra with many overlapping resonances per
gauss \cite{APetrov:12,AFrisch:14,KBaumann:14}.  This makes it impractical to
assign quantum labels to individual resonances and may be a challenge to
interaction ``tailoring'' and molecule formation.

In this paper, we study magnetic s-wave Feshbach resonances in binary mixtures
of ground-state Li atoms and bosonic Er isotopes.  The Li+Er system may be
specially appealing for ultracold experiments in optical lattices: Dipolar species
with tunable interactions are key to studying the effects of long-range
anisotropies, quantum magnetism, disorder and quantum collective behavior
\cite{MHGdeMiranda:11a,GQuemener:12,BYan:13,ADePaz:13,KAikawa:14b}.
Very importantly, such Feshbach resonances may be used for magnetoassociation of
LiEr molecules, starting from ground-state atoms in order to avoid limiting 
background losses \cite{MLGonzalez-Martinez:13a}.  Ground-state LiEr molecules 
have both magnetic and electric dipole moments
\footnote{Preliminary multireference configuration interaction (MRCI) calculations
predict a permanent electric dipole moment of about 1.5~Debye for the lowest
$^4\Sigma$ state, and the magnetic dipole moment may be up to approximately 8 Bohr
magnetons.},
and may be controlled with applied electric and magnetic fields which further
enhances their applicability
\cite{MABaranov:12,GQuemener:12,MTomza:13b,MTomza:14a}.  In addition, Er is a 
heavy atom thus ultracold LiEr may be used to study the time variation of 
fundamental \textit{constants} \cite{JJHudson:02,TZelevinsky:08}, while the 
extreme mass imbalance in the system makes it specially well-suited for exploring 
Efimov physics \cite{RPires:14a}.

We carried out coupled-channel calculations using the theory in
Ref.~\cite{MLGonzalez-Martinez:13c}.  The Hamiltonian can be written
\begin{equation}
 \hat{\mathcal{H}} = - \frac{\hbar^2}{2\mu} R^{-1} \frac{d^2}{dR^2} R
                     + \frac{\hbar^2\hat{L}^2}{2\mu R^2}
                     + \hat{\mathcal{H}}_\mathrm{Li}
                     + \hat{\mathcal{H}}_\mathrm{Er}
                     + \hat{\mathcal{U}},
 \label{eq:Heff}
\end{equation}
where $\mu$ is the reduced mass for the collision, $R$ is the interatomic
distance and $\hat{L}$ is the space-fixed operator for the end-over-end
rotation.  $\hat{\mathcal{H}}_\mathrm{Li}$ and $\hat{\mathcal{H}}_\mathrm{Er}$
describe the isolated atoms and are taken to be
\begin{eqnarray}
 \hat{\mathcal{H}}_\mathrm{Li} &=&
  b_\mathrm{F, Li} \hat{\imath}_\mathrm{Li}\cdot \hat{s}_\mathrm{Li} +
  (g_S \mu_\mathrm{B} \hat{s}_\mathrm{Li} - g_{i\mathrm{Li}} \mu_\mathrm{N}
  \hat{\imath}_\mathrm{Li}) \cdot \boldsymbol{B}; \nonumber \\
 \hat{\mathcal{H}}_\mathrm{Er} &=&
  a^\mathrm{so}_\mathrm{Er} \hat{l}_\mathrm{Er} \cdot \hat{s}_\mathrm{Er}
  + (g'_L \mu_\mathrm{B} \hat{l}_\mathrm{Er}
  + g_S \mu_\mathrm{B} \hat{s}_\mathrm{Er}) \cdot \boldsymbol{B}.
 \label{eq:H1+2}
\end{eqnarray}
Here, $\hat{\imath}_\mathrm{Li}$ and $\hat{s}_\mathrm{Li}$ are the Li nuclear
and electronic spin operators, while $\hat{l}_\mathrm{Er}$ and
$\hat{s}_\mathrm{Er}$ denote the Er electronic orbital and spin operators (all
bosonic Er isotopes have zero nuclear spin); $\boldsymbol{B}$ is the external
magnetic field.  $g_S \approx 2$, $g_{i\mathrm{Li}}$ and $g'_L \approx 1$ are
the electron, Li nuclear and orbital $g$ factors, while $\mu_\mathrm{B}$ and
$\mu_\mathrm{N}$ are the Bohr and nuclear magnetons.  The hyperfine coupling
constants for $^{6,7}$Li ($b_\mathrm{F, Li}$) and the nuclear $g$ factors were
taken from Refs.~\cite{ABeckmann:74,NJStone:05}.  The spin-orbit coupling
constant for Er, $a^\mathrm{so}_\mathrm{Er} = -1159.7215 \times hc$~cm$^{-1}$,
was calculated from the splitting between the two lowest Er states, $^3$H$_6$ and
$^3$H$_5$ \cite{WCMartin:78}, assuming Russel-Saunders coupling.
$\hat{\mathcal{U}}$ describes all interactions between the atoms and includes
the electronic potential $\hat{V}$ and the direct dipolar interaction between
the atoms magnetic moments $\hat{\mathcal{H}}_\mathrm{dip}$
\cite{MLGonzalez-Martinez:13c}.  Following Krems \textit{et al.}\
\cite{RVKrems:04b}, we decompose $\hat{V}$ into functions with well-defined
total spin $S$ and space-fixed spin projection $M_S$, which are then expanded in
Legendre polynomials---assuming that $l_\mathrm{Er} = 5$ is conserved at all
values of $R$.  The expansion coefficients, $\hat{V}^S_k(R)$ ($k = 0, 2,\ldots,
2l_\mathrm{Er}$), are linear combinations of the Born-Oppenheimer potentials
\cite{RVKrems:04b,MLGonzalez-Martinez:13c}; these can be split into isotropic
$V^S_0$ and anisotropic $V^S_{k\ne0}$ terms, the latter depending only on the
energy \emph{differences} between Born-Oppenheimer states.

The interaction between Li($^2$S) and Er($^3$H) gives rise to twelve electronic
states: six states corresponding to $|\Lambda| = 0,\ldots,l_\mathrm{Er}$---the 
absolute value of the projection of the electronic orbital angular momentum onto 
the interatomic axis---for each $S$, $S_-=\frac{1}{2}$ and $S_+ = \frac{3}{2}$.
We calculated the short-range interaction energies for all Li($^2$S)+Er($^3$H)
states using the complete active space self-consistent field method (CASSCF)
implemented in \textsc{molpro} \cite{MOLPRO2012_brief}.  The active space
includes $1s2s$ orbitals for Li and $4f6s6p$ orbitals for Er.  We used the
high-quality uncontracted aug-cc-pVQZ basis sets by Prascher \textit{et al.}\
\cite{BPPrascher:11} for the Li atom.  For Er, we used the quasirelativistic
effective core potential by Dolg \textit{et al.}\ \cite{MDolg:89} (ECP28MWB) for
the first 28 electrons, with uncontracted $s$ and $p$ shells, to which we added
$h$ functions with exponent 0.45.  The basis was augmented with extra diffused
functions using the even-tempered scheme in \textsc{molpro}
\cite{MOLPRO2012_brief}.  The Li+Er states lie relatively close in energy and
special care is needed to avoid their mixing.  We start our \textit{ab initio}
calculations at large $R = 50~a_0$, calculating the starting orbitals by merging 
those of the isolated atoms; this way, we obtain properly 22-fold degenerate
states for Li+Er.  In each following calculation, we take a step inwards in $R$
and use the orbitals converged from the precedent geometry as starting orbitals.
We obtain potentials corresponding to pure $\Lambda=0,\ldots \pm l_\mathrm{Er}$
states by controlling the $\Lambda$ quantum number of the molecule.

With this scheme, we get realistic short-range energy differences between all
Li+Er potentials, and thus realistic anisotropies $V^S_{k\ne0}$, from our CASSCF
calculations.  The active space, however, is not sufficiently large for describing
the Li+Er dispersion accurately, which mainly affects the $V^S_0$ isotropic terms.
This can be fixed by noting that the dispersion interaction in analogous systems
is relatively spin-independent, with the spin-dependent exchange-dispersion
energy being typically very small \cite{BJeziorski:94}.
We further assumed that the spin-independent dispersion interaction in Li+Er is
similar to that in Li+Yb, which is supported by the similarity of their isotropic 
dispersion coefficients ($1\,594$~a.u.\ for Li+Yb \cite{PZhang:10} and
$1\,508$~a.u.\ for Li+Er---see calculation details below).  We used the Heisenberg 
spin-exchange  model \cite{WHeisenberg:28,AABuchachenko:09,TVTscherbul:10a} for 
the isotropic  potentials, $V^S_0(R) = V_0(R)-2J_0(R) \hat{s}_\mathrm{Li} \cdot
\hat{s}_\mathrm{Er}$, and replaced the spin-independent term $V_0$ with the
LiYb potential obtained by Zhang and coworkers using the best \textit{ab initio}
methods available \cite{PZhang:10}.  $J_0 = -[V^{S_+}_0 - V^{S_-}_0]/2S_+$
depends on energy differences only and is thus recovered from our CASSCF
calculations.  We inter- and extrapolated all $V^S_k$ curves with the reproducing 
kernel Hilbert space (RKHS) approach of Ho and Rabitz \cite{T-SHo:96a}.  Both
$V^S_0$ isotropic potentials were constrained at long range to have $C_{6,0} =
1\,508$~a.u., calculated from Tang's combination rule \cite{KTTang:69} with the
values of the static polarizability and dispersion coefficients for Li$_2$
\cite{ADerevianko:10} and Er$_2$ \cite{MLepers:14a}.  The $V^S_2$ anisotropies
were constrained to have $C_{6,2} = 35.04$~a.u., obtained from 
$C_{6,0}$ and the anisotropic and isotropic polarizabilities of Er (\textit{cf.},
Ref.~\cite{XChu:07}).  We neglected the Van der Waals expansion in $V^S_4$ and
higher-order terms as they decay faster than $R^{-6}$.

\begin{figure}[!t]
 \includegraphics[width=86mm]{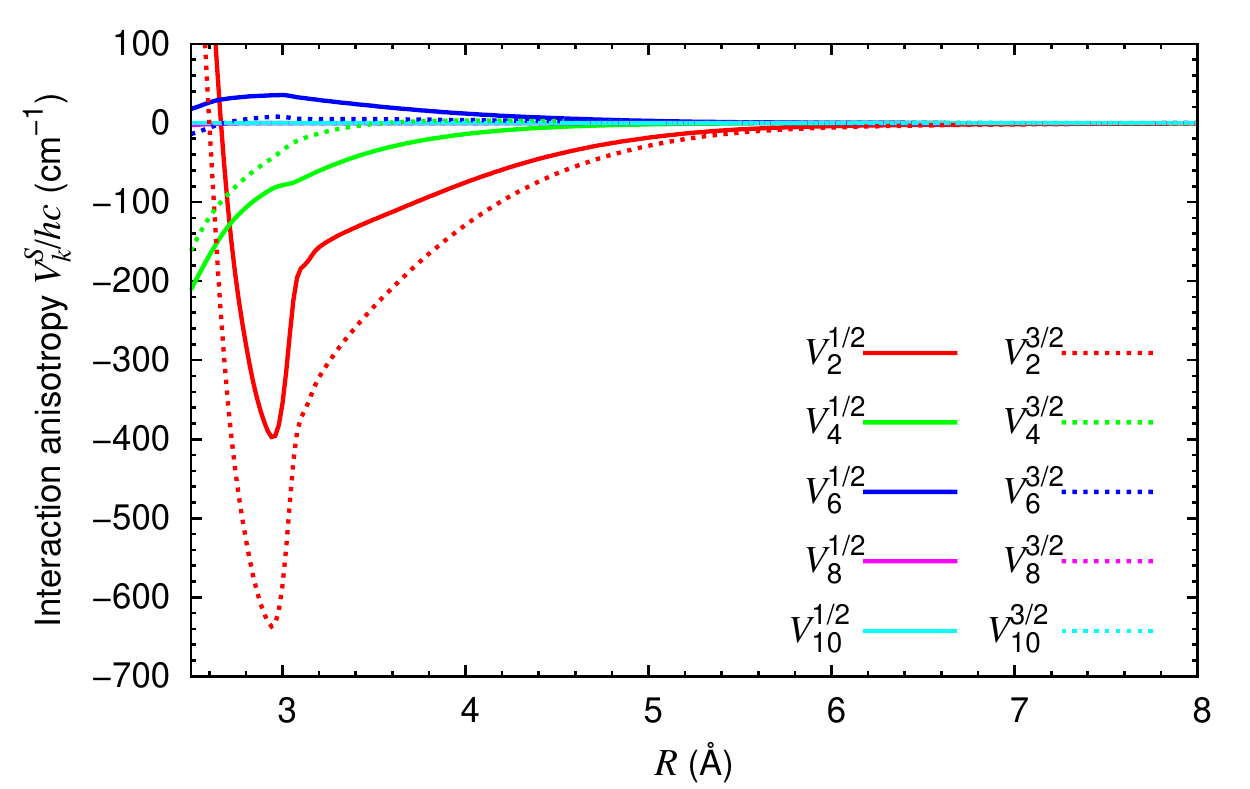}
 \caption{(Color online) Calculated interaction anisotropies in Li+Er for low
  (solid) and high (dotted) total electronic spin.}
 \label{fig:VSk}
\end{figure}
Figure~\ref{fig:VSk} shows the CASSCF interaction anisotropies $V^S_{k\ne0}$ for
Li+Er.  The magnitude of the spin-exchange interaction potential curves is 
qualitatively similar to recent calculations by Tomza on Li+Eu \cite{MTomza:14a}, 
for which the spin-exchange interaction near equilibrium was about 600~cm$^{-1}$. 
The fact that spin-exchange in this system is much smaller than for alkali dimers 
is explained by suppression due to the outermost $6s^2$ shell of Er.  The same
mechanism reduces the anisotropies related to the electronic orbital angular 
momentum: they are on the order of few hundreds of cm$^{-1}$ near the Van der 
Waals minimum for $V^S_2$ terms and orders of magnitude lower for higher-order 
anisotropies.  The mechanism of suppression of $L$-anisotropy was found earlier by
Krems and coworkers for the He+transition metal systems \cite{RVKrems:05b} and for 
Yb+Tm by Buchachenko \textit{et al.}\ \cite{AABuchachenko:07}.
This is in contrast to the Li+Yb($6s^16p^1$) system where the $L$-anisotropy is
on the order of thousands of cm$^{-1}$ \cite{MLGonzalez-Martinez:13a}.

We studied s-wave magnetic Feshbach resonances in the Li+Er systems using the
\textsc{molscat} \cite{JMHutson:MOLSCAT14,MLGonzalez-Martinez:07a} and \textsc{field} \cite{JMHutson:FIELD1} packages.
We used computational methods analogous to those in previous work on Li+Yb 
\cite{MLGonzalez-Martinez:13a} and H+F \cite{MLGonzalez-Martinez:13c}.  The 
collision energy in our scattering calculations is fixed at
$1\times k_\mathrm{B}$~nK.  Convergence in the partial-wave expansion was
achieved with $L_\mathrm{max} = 10$.  We consider calculations with both atoms in 
their lowest Zeeman state.

\begin{figure}[!t]
 \includegraphics[width=86mm]{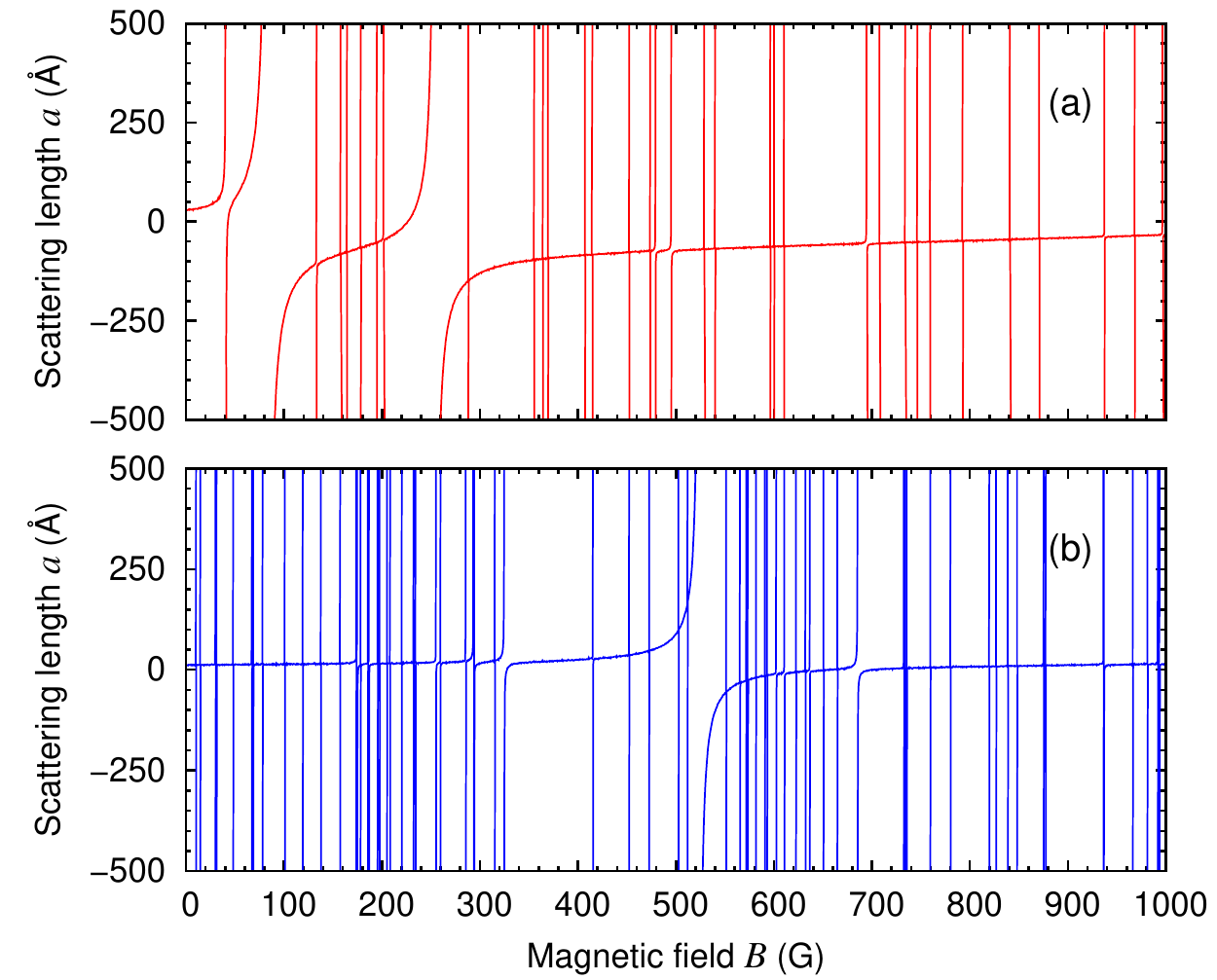}
 \caption{(Color online) Magnetic field dependence of the scattering length
  $a(B)$ for ground-state collisions of $^{166}$Er with: (a) $^6$Li, and (b)
  $^7$Li.}
 \label{fig:avsB_Li2S+166Er3H}
\end{figure}
Figure~\ref{fig:avsB_Li2S+166Er3H} shows the predicted magnetic field dependence
of the Li+$^{166}$Er scattering length, depicting many Feshbach resonances of
widths between 0.1 and 50~G that are very promising candidates for precise tuning 
and magnetoassociation.  These resonances are immune to background losses since 
both species interact in their absolute ground state.  We located 35/69 resonances 
below $1\,000$~G for $^6$Li/$^7$Li+$^{166}$Er, and calculated mean densities of
$\bar{\rho}_\mathrm{^6Li} = 0.0364$~G$^{-1}$ and
$\bar{\rho}_\mathrm{^7Li} = 0.0730$~G$^{-1}$: The combination of light and 
heavy species yields a much wider rovibrational spacing than for a heavy+heavy 
system, hence Li+Er Feshbach spectra are much less congested than others involving 
highly-magnetic atoms \cite{APetrov:12,AFrisch:14,KBaumann:14}.

The couplings responsible for these resonances have been studied for analogous 
systems, \textit{cf.},
Refs.~\cite{RVKrems:04b,MLGonzalez-Martinez:13a,MLGonzalez-Martinez:13c}.
These involve orbital- and/or spin-anisotropies from the electronic potentials,
with the widest resonances due to states dominated by low $L$ quantum numbers.
The general trends in our calculations support this reasoning.  The widest 
resonances in Fig.~\ref{fig:avsB_Li2S+166Er3H} ($\Delta_B > 10$~G) correspond to 
states with over 55\% and up to 45\% compositions from $L = 0$ and 2, 
respectively, which are mainly coupled to the initial state by $V^S_{k = 0, 2}$ 
terms.  An intermediate resonance ``class'' ($1 < \Delta_B < 5$~G) arises from 
states with over 30\% composition of $L = 4$, and less than 15\% of $L = 0$,
involving $V^S_{4}$ coupling terms.  Narrower resonances are due to states with 
non-negligible higher-order orbital excitations of the complex, $L \ge 6$.
Removing high-order anisotropies $V^S_{k > 4}$, however, does not affect the
overall resonance pattern since indirect mechanisms involving $V^S_{k \le 4}$
terms also couple the resonant states to the continuum.
The contribution from dipole-dipole interactions is very small compared to
that of the potential terms and was found to be negligible, as for Li+Yb
\cite{MLGonzalez-Martinez:13a}.

\begin{figure}[!t]
 \includegraphics[width=86mm]{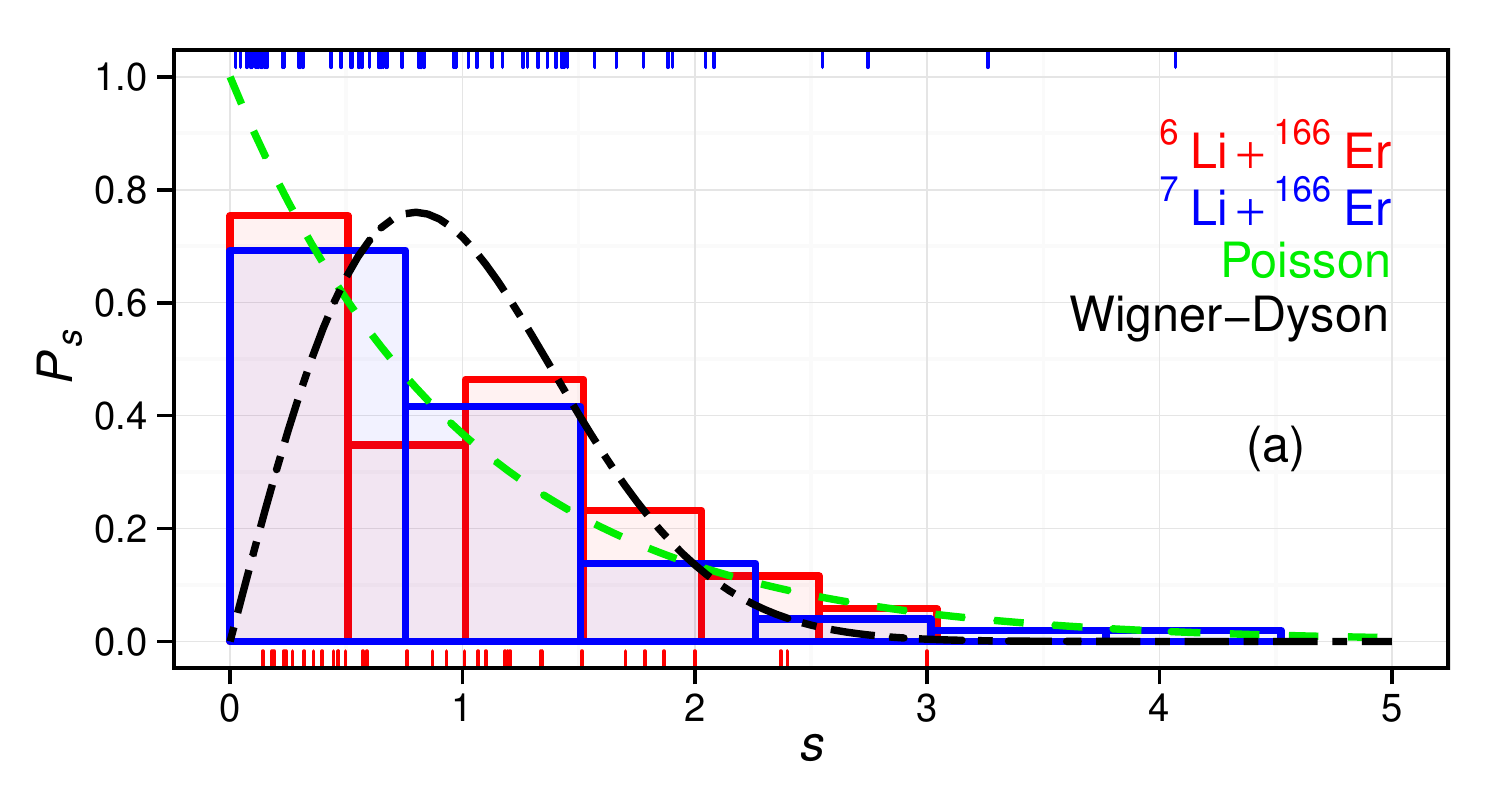}
 \includegraphics[width=86mm]{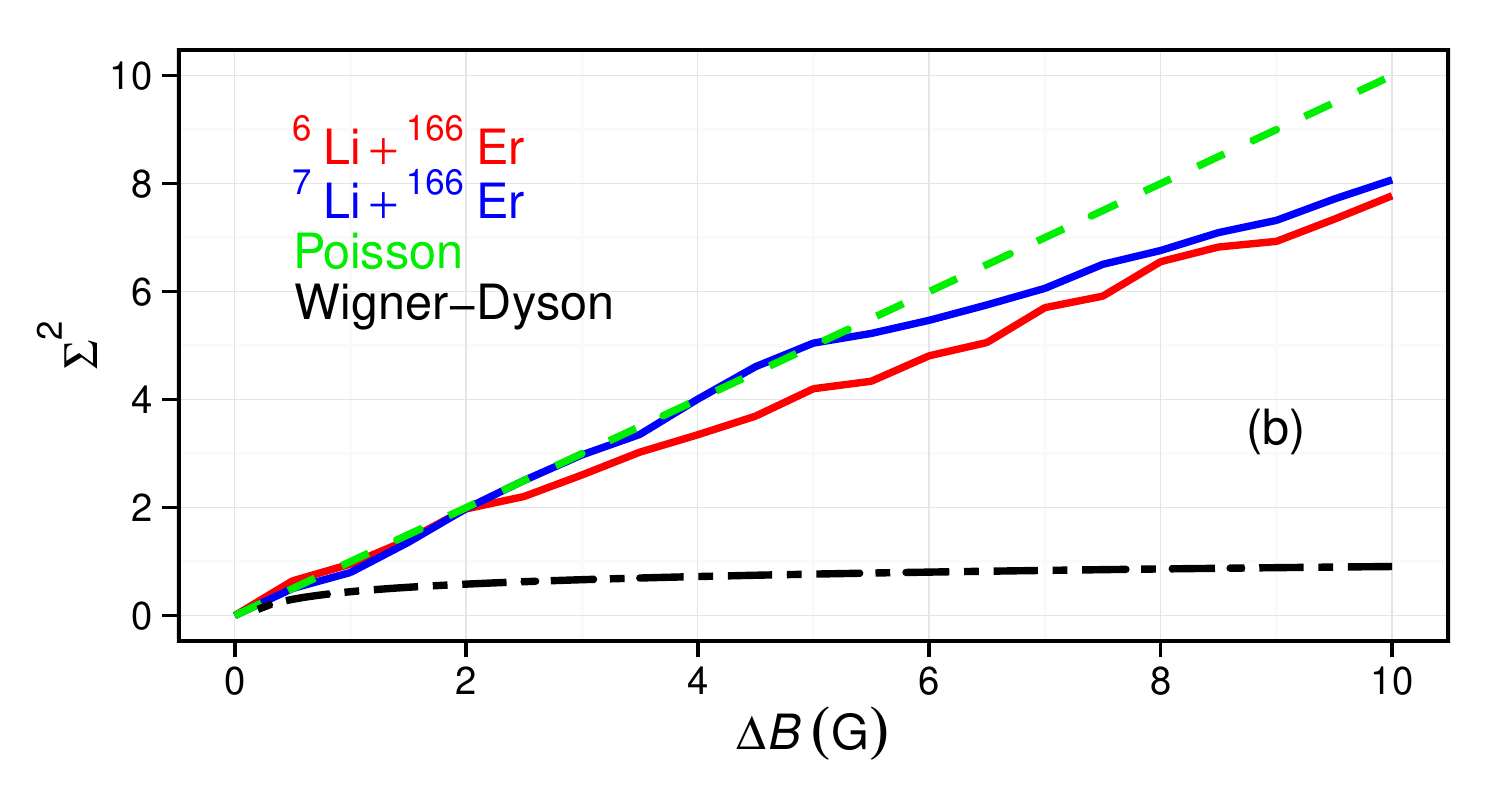}
 \caption{(Color online) (a) Nearest-neighbor spacing distributions, and
  (b) number variances for the $^6$Li/$^7$Li+$^{166}$Er (solid red/blue) spectra.
  Poisson (dashed, green) and Wigner-Dyson (two-dashed, black) curves are added
  for comparison.}
 \label{fig:stats}
\end{figure}

\begin{table*}[!t]
 \begin{center}
  \caption{Summary of statistical tests and/or model fitting to Li+$^{166}$Er
   Feshbach spectra ($\tilde{\chi}$ and $\tilde{G}$ are \textit{reduced}
   quantities).
  \label{tab:stats}}
  \begin{tabular}{ccccccccccccc}
   \hline\hline
     Lithium        && \multicolumn{2}{c}{$\chi^2$-test} && \multicolumn{2}{c}{$G$-test} && Bayes' LLR   && Two-gap     && Fit to Brody \\ \cline{3-4} \cline{6-7}
     isotope        && Poisson & Wigner-Dyson            && Poisson & Wigner-Dyson      && $\log{[P(\boldsymbol{x}|\mathrm{H}_\mathrm{P})/P(\boldsymbol{x}|\mathrm{H}_\mathrm{WD})]}$ && correlation && distribution \\ \hline
   $^6$Li && $\tilde{\chi}^2 = 0.760$ & $\tilde{\chi}^2 = 2.53$    && $\tilde{G} = 0.982$ & $\tilde{G} = 2.08$ && 3.30 && $\eta_r = 0.306$ && $\eta = 0.0738$   \\
             && ($p = 0.582$)            & ($p = 0.0435$)             && ($p = 0.436$)            & ($p = 0.0518$)     &&      &&                  && ($\mathrm{MSE} = 0.0101$)  \\
   $^7$Li && $\tilde{\chi}^2 = 1.40$  & $\tilde{\chi}^2 > 10^{10}$ && $\tilde{G} = 0.760$ & $\tilde{G} = 8.06$ && 32.9 && $\eta_r = 0.161$ && $\eta = 0.241$    \\
             && ($p = 0.246$)            & ($p < 10^{-3}$)            && ($p = 0.654$)            & ($p < 10^{-11}$)   &&      &&                  && ($\mathrm{MSE} = 0.000379$) \\
   \hline\hline
  \end{tabular}
 \end{center}
\end{table*}

Studying the statistics of the calculated Feshbach spectra is key to unraveling
their most general and robust properties.  We performed an analysis similar to
that of Frisch and coworkers \cite{AFrisch:14}, with added $G$-tests and Bayes'
logarithmic likelihood ratio (LLR) calculations.  Figure~\ref{fig:stats} shows the
nearest-neighbour spacing (NNS) distribution and the number variance for the
calculated $^{6,7}$Li+$^{166}$Er Feshbach spectra---predictions from the Poisson
and Wigner-Dyson models are added for comparison.  Table~\ref{tab:stats}
summarizes all statistical tests performed and details on the fits to Brody NNS
distributions.  All our tests indicate that Li+Er Feshbach spectra arise from
weakly-interacting levels exhibiting very low spectra rigidity and level
repulsion, with Poisson-like models providing a significantly better description
of the spectra.  These are all characteristics of non-chaotic spectra and make it
possible to identify and tune selected individual Feshbach resonances for
interaction tailoring and/or magnetoassociation.

Erbium has five bosonic isotopes and the change in Li+Er reduced mass with
respect to that of Li+$^{166}$Er is about $\pm 0.1$\% for $^{6,7}$Li.
The small changes in $\mu$, together with the non-chaotic nature and
relatively low densities of the spectra all support and would simplify
predicting the Feshbach spectra for different Er isotopes.
If $\delta E_v$ is the isotopic shift in the binding energy of a
near-dissociation molecular state and $\delta\mu_\mathrm{res}$ is the difference
between the magnetic moments of the molecule and free atoms (at resonance), the 
isotopic shift in resonance position,
$\delta B_\mathrm{res} \approx \delta E_v/\delta\mu_\mathrm{res}$, may be
estimated for an $R^{-6}$ potential to be \cite{RJLeRoy:70a,GFGribakin:93}
\begin{equation}
 \delta B_\mathrm{res} \approx
  \frac{3 H^3_{6,1}}{2 \delta\mu_\mathrm{res}}
  \left( v^\mathrm{WKB}_{\mathrm{D},1} - v \right)^2
  \left( v + \frac{1}{2} \right) \left(1 - \frac{\mu_2}{\mu_1} \right),
 \label{eq:dBres}
\end{equation}
where we approximated
$\delta E_v \approx (\mathrm{d}E_v/\mathrm{d}\mu)|_{\mu_1} (\mu_2 - \mu_1)$.
Both the parameter
$H_{6,1} \approx 3.4346\,\hbar \left(\mu_1 C^{1/3}_{6,0}\right)^{-1/2}$ and 
the WKB noninteger quantum number at dissociation,
$v^\mathrm{WKB}_\mathrm{D,1} = (1/\pi)\arctan{(1 - a_\mathrm{bg,1}/\bar{a}_1) -
3/8}$---with
$\bar{a}_1 \approx 0.47799 \left(2 \mu_1 C_{6,0}/\hbar^2 \right)^{1/4}$---are 
calculated for a reference system.
As expected, Eq.~\eqref{eq:dBres} shows that $\delta B_\mathrm{res} > 0$ if
$\mu_2 > \mu_1$, and viceversa ($\delta\mu_\mathrm{res} < 0$ in our case).
Eq.~\eqref{eq:dBres} suggests that it may be possible to estimate the position of 
an equivalent resonance in a different isotopologue, while experiments may extract
key information on background scattering lengths, quantum numbers and
non-adiabatic effects from isotopic resonance shifts.

It is not yet possible to calculate \textit{quantitatively} correct
\textit{ab-initio} Li+Er potentials, hence the robustness of our conclusions need 
to be assessed for their possible dependence on these.  The energy of the top 
vibrational state may be estimated from the Van der Waals coefficient $C_{6,0}$
and reduced mass $\mu$ \cite{BGao:04} to lie between 0 and
$13\hbar^2/(\mu R^2_\mathrm{vdW}) \approx 6.3 \times h$~GHz---where the 
characteristic length scale,
$R_\mathrm{vdW} = \frac{1}{2}(2\mu C_{6,0}/\hbar^2)^{1/4}$.
The smallest difference between the magnetic moments of a scattering
state and an $L = 0$-supported bound state is $\mu_\mathrm{B}$.  The largest
magnetic field at which resonances from such levels might occur is thus
$6.3 \times h~\mathrm{GHz}/\mu_\mathrm{B} \approx 5\,000$~G.  Resonances resulting 
from larger magnetic moment differences may appear at much lower fields, however,
$5\,000$~G is the smallest relevant range to study the dependence of our results
on the interaction potentials.

\begin{figure}[!t]
 \includegraphics[width=88mm]{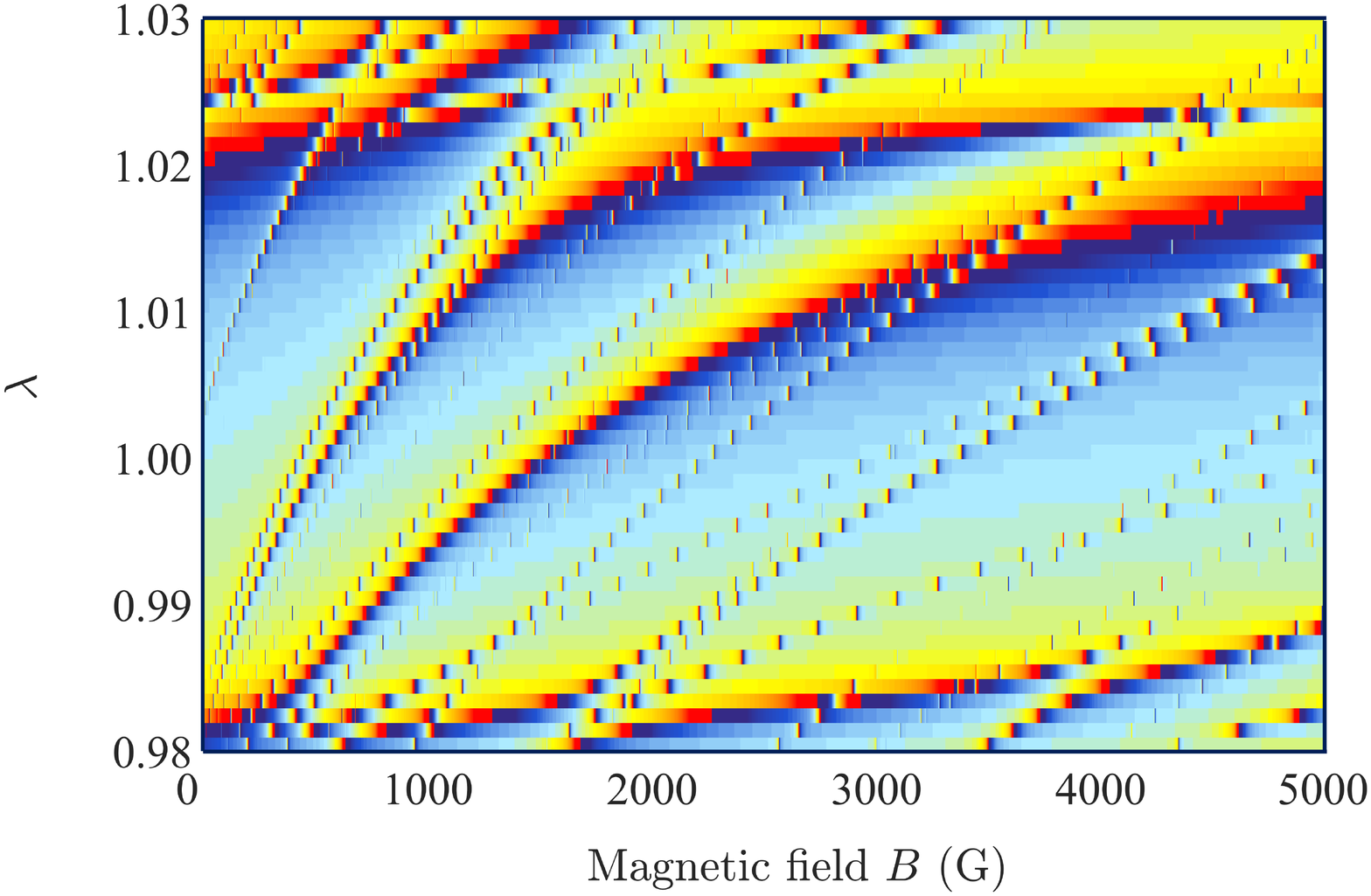}
 \caption{(Color online).  Contour plot of the scattering length $a$ for
  $^7$Li+$^{166}$Er as a function of the magnetic field and a scaling parameter.
  Red/blue indicates the highest/lowest values.}
 \label{fig:lscan}
\end{figure}
Figure~\ref{fig:lscan} shows a contour plot of the scattering length for
$^7$Li+$^{166}$Er as a function of the magnetic field, $B$, and a parameter
$\lambda$ used to scale the isotropic potentials, $V'^S_0 = \lambda V^S_0$.  To 
keep a practical computational cost, the calculations neglect hyperfine 
terms and use $L_\mathrm{max} = 6$.  The approximations preserve the main resonant 
features---compare with Fig.~\ref{fig:avsB_Li2S+166Er3H}(b)---while the 
broadest resonances ($\Delta_B > 5$~G) are most apparent at the grid's resolution.
Fig.~\ref{fig:lscan} reveals that there always exist broad magnetic Feshbach 
resonances at relatively low fields.
Additionally, it shows that $\lambda = 1$ is statistically representative for a 
much wider $\lambda$ range, except around 0.985 and 1.025 for which the background 
scattering length becomes very large and enhances all resonance widths.

This work provides robust theoretical evidence that low-field magnetic Feshbach
resonances immune to background losses exist for Li+Er, with widths
$\Delta_B > 0.1$ well within current experimental resolution.
Li+Er spectra are predicted to be non-chaotic, while remaining conveniently dense,
in contrast with other systems involving highly-magnetic atoms such as Er+Er 
\cite{AFrisch:14}.
The predicted resonances may be resolved and independently addressed thus
opening the door for precise tuning of Li+Er interactions and/or 
magnetoassociation into LiEr molecules.
The characteristics of the spectra make it also possible to predict resonance 
positions for different isotopologues from measurements on
a reference system, which would greatly simplify experiments with various Er 
isotopes and give key insight into non-adiabatic effects.

M.L.G.M.\ acknowledges funding from the European Community's Seventh Framework
Programme (FP7/2007-2013) under Grant Agreement No.~330623.
P.S.\.{Z}. acknowledges funding from the Polish National Science Center (NCN)
grant DEC-2012/07/B/ST2/00235.


\begin{thebibliography}{67}
\expandafter\ifx\csname natexlab\endcsname\relax\def\natexlab#1{#1}\fi
\expandafter\ifx\csname bibnamefont\endcsname\relax
  \def\bibnamefont#1{#1}\fi
\expandafter\ifx\csname bibfnamefont\endcsname\relax
  \def\bibfnamefont#1{#1}\fi
\expandafter\ifx\csname citenamefont\endcsname\relax
  \def\citenamefont#1{#1}\fi
\expandafter\ifx\csname url\endcsname\relax
  \def\url#1{\texttt{#1}}\fi
\expandafter\ifx\csname urlprefix\endcsname\relax\def\urlprefix{URL }\fi
\providecommand{\bibinfo}[2]{#2}
\providecommand{\eprint}[2][]{\url{#2}}

\bibitem[{\citenamefont{Carr et~al.}(2009)\citenamefont{Carr, DeMille, Krems,
  and Ye}}]{LDCarr:09b}
\bibinfo{author}{\bibfnamefont{L.~D.} \bibnamefont{Carr}},
  \bibinfo{author}{\bibfnamefont{D.}~\bibnamefont{DeMille}},
  \bibinfo{author}{\bibfnamefont{R.~V.} \bibnamefont{Krems}}, \bibnamefont{and}
  \bibinfo{author}{\bibfnamefont{J.}~\bibnamefont{Ye}}, \bibinfo{journal}{New
  J. Phys.} \textbf{\bibinfo{volume}{11}}, \bibinfo{pages}{055049}
  (\bibinfo{year}{2009}).

\bibitem[{\citenamefont{Dulieu and Gabbanini}(2009)}]{ODulieu:09a}
\bibinfo{author}{\bibfnamefont{O.}~\bibnamefont{Dulieu}} \bibnamefont{and}
  \bibinfo{author}{\bibfnamefont{C.}~\bibnamefont{Gabbanini}},
  \bibinfo{journal}{Rep. Prog. Phys.} \textbf{\bibinfo{volume}{72}},
  \bibinfo{pages}{086401} (\bibinfo{year}{2009}).

\bibitem[{\citenamefont{Bloch et~al.}(2012)\citenamefont{Bloch, Dalibard, and
  Nascimb\`ene}}]{IBloch:12a}
\bibinfo{author}{\bibfnamefont{I.}~\bibnamefont{Bloch}},
  \bibinfo{author}{\bibfnamefont{J.}~\bibnamefont{Dalibard}}, \bibnamefont{and}
  \bibinfo{author}{\bibfnamefont{S.}~\bibnamefont{Nascimb\`ene}},
  \bibinfo{journal}{Nat. Phys.} \textbf{\bibinfo{volume}{8}},
  \bibinfo{pages}{267} (\bibinfo{year}{2012}).

\bibitem[{\citenamefont{Baranov et~al.}(2012)\citenamefont{Baranov, Dalmonte,
  Pupillo, and Zoller}}]{MABaranov:12}
\bibinfo{author}{\bibfnamefont{M.~A.} \bibnamefont{Baranov}},
  \bibinfo{author}{\bibfnamefont{M.}~\bibnamefont{Dalmonte}},
  \bibinfo{author}{\bibfnamefont{G.}~\bibnamefont{Pupillo}}, \bibnamefont{and}
  \bibinfo{author}{\bibfnamefont{P.}~\bibnamefont{Zoller}},
  \bibinfo{journal}{Chem. Rev.} \textbf{\bibinfo{volume}{112}},
  \bibinfo{pages}{5012} (\bibinfo{year}{2012}).

\bibitem[{\citenamefont{Micheli et~al.}(2006)\citenamefont{Micheli, Brennen,
  and Zoller}}]{AMicheli:06}
\bibinfo{author}{\bibfnamefont{A.}~\bibnamefont{Micheli}},
  \bibinfo{author}{\bibfnamefont{G.~K.} \bibnamefont{Brennen}},
  \bibnamefont{and} \bibinfo{author}{\bibfnamefont{P.}~\bibnamefont{Zoller}},
  \bibinfo{journal}{Nat. Phys.} \textbf{\bibinfo{volume}{2}},
  \bibinfo{pages}{341} (\bibinfo{year}{2006}).

\bibitem[{\citenamefont{Micheli et~al.}(2007)\citenamefont{Micheli, Pupillo,
  B\"uchler, and Zoller}}]{AMicheli:07}
\bibinfo{author}{\bibfnamefont{A.}~\bibnamefont{Micheli}},
  \bibinfo{author}{\bibfnamefont{G.}~\bibnamefont{Pupillo}},
  \bibinfo{author}{\bibfnamefont{H.~P.} \bibnamefont{B\"uchler}},
  \bibnamefont{and} \bibinfo{author}{\bibfnamefont{P.}~\bibnamefont{Zoller}},
  \bibinfo{journal}{Phys. Rev. A} \textbf{\bibinfo{volume}{76}},
  \bibinfo{pages}{043604} (\bibinfo{year}{2007}).

\bibitem[{\citenamefont{B\"uchler et~al.}(2007)\citenamefont{B\"uchler, Demler,
  Lukin, Micheli, Prokof'ev, Pupillo, and Zoller}}]{HPBuchler:07}
\bibinfo{author}{\bibfnamefont{H.~P.} \bibnamefont{B\"uchler}},
  \bibinfo{author}{\bibfnamefont{E.}~\bibnamefont{Demler}},
  \bibinfo{author}{\bibfnamefont{M.}~\bibnamefont{Lukin}},
  \bibinfo{author}{\bibfnamefont{A.}~\bibnamefont{Micheli}},
  \bibinfo{author}{\bibfnamefont{N.}~\bibnamefont{Prokof'ev}},
  \bibinfo{author}{\bibfnamefont{G.}~\bibnamefont{Pupillo}}, \bibnamefont{and}
  \bibinfo{author}{\bibfnamefont{P.}~\bibnamefont{Zoller}},
  \bibinfo{journal}{Phys. Rev. Lett.} \textbf{\bibinfo{volume}{98}},
  \bibinfo{pages}{060404} (\bibinfo{year}{2007}).

\bibitem[{\citenamefont{Ni et~al.}(2010)\citenamefont{Ni, Ospelkaus, Wang,
  Qu\'em\'ener, Neyenhuis, {de Miranda}, Bohn, Ye, and Jin}}]{K-KNi:10}
\bibinfo{author}{\bibfnamefont{K.-K.} \bibnamefont{Ni}},
  \bibinfo{author}{\bibfnamefont{S.}~\bibnamefont{Ospelkaus}},
  \bibinfo{author}{\bibfnamefont{D.}~\bibnamefont{Wang}},
  \bibinfo{author}{\bibfnamefont{G.}~\bibnamefont{Qu\'em\'ener}},
  \bibinfo{author}{\bibfnamefont{B.}~\bibnamefont{Neyenhuis}},
  \bibinfo{author}{\bibfnamefont{M.~H.~G.} \bibnamefont{{de Miranda}}},
  \bibinfo{author}{\bibfnamefont{J.~L.} \bibnamefont{Bohn}},
  \bibinfo{author}{\bibfnamefont{J.}~\bibnamefont{Ye}}, \bibnamefont{and}
  \bibinfo{author}{\bibfnamefont{D.~S.} \bibnamefont{Jin}},
  \bibinfo{journal}{Nature} \textbf{\bibinfo{volume}{464}},
  \bibinfo{pages}{1324} (\bibinfo{year}{2010}).

\bibitem[{\citenamefont{Ospelkaus et~al.}(2010)\citenamefont{Ospelkaus, Ni,
  Wang, de~Miranda, Neyenhuis, Qu\'em\'ener, Julienne, Bohn, Jin, and
  Ye}}]{SOspelkaus:10b}
\bibinfo{author}{\bibfnamefont{S.}~\bibnamefont{Ospelkaus}},
  \bibinfo{author}{\bibfnamefont{K.-K.} \bibnamefont{Ni}},
  \bibinfo{author}{\bibfnamefont{D.}~\bibnamefont{Wang}},
  \bibinfo{author}{\bibfnamefont{M.~H.~G.} \bibnamefont{de~Miranda}},
  \bibinfo{author}{\bibfnamefont{B.}~\bibnamefont{Neyenhuis}},
  \bibinfo{author}{\bibfnamefont{G.}~\bibnamefont{Qu\'em\'ener}},
  \bibinfo{author}{\bibfnamefont{P.~S.} \bibnamefont{Julienne}},
  \bibinfo{author}{\bibfnamefont{J.~L.} \bibnamefont{Bohn}},
  \bibinfo{author}{\bibfnamefont{D.~S.} \bibnamefont{Jin}}, \bibnamefont{and}
  \bibinfo{author}{\bibfnamefont{J.}~\bibnamefont{Ye}},
  \bibinfo{journal}{Science} \textbf{\bibinfo{volume}{327}},
  \bibinfo{pages}{853} (\bibinfo{year}{2010}).

\bibitem[{\citenamefont{{de Miranda} et~al.}(2011)\citenamefont{{de Miranda},
  Chotia, Neyenhuis, Wang, Qu\'em\'ener, Ospelkaus, Bohn, Ye, and
  Jin}}]{MHGdeMiranda:11a}
\bibinfo{author}{\bibfnamefont{M.~H.~G.} \bibnamefont{{de Miranda}}},
  \bibinfo{author}{\bibfnamefont{A.}~\bibnamefont{Chotia}},
  \bibinfo{author}{\bibfnamefont{B.}~\bibnamefont{Neyenhuis}},
  \bibinfo{author}{\bibfnamefont{D.}~\bibnamefont{Wang}},
  \bibinfo{author}{\bibfnamefont{G.}~\bibnamefont{Qu\'em\'ener}},
  \bibinfo{author}{\bibfnamefont{S.}~\bibnamefont{Ospelkaus}},
  \bibinfo{author}{\bibfnamefont{J.~L.} \bibnamefont{Bohn}},
  \bibinfo{author}{\bibfnamefont{J.}~\bibnamefont{Ye}}, \bibnamefont{and}
  \bibinfo{author}{\bibfnamefont{D.~S.} \bibnamefont{Jin}},
  \bibinfo{journal}{Nat. Phys.} \textbf{\bibinfo{volume}{7}},
  \bibinfo{pages}{502} (\bibinfo{year}{2011}).

\bibitem[{\citenamefont{Derevianko and Cannon}(2004)}]{ADerevianko:04}
\bibinfo{author}{\bibfnamefont{A.}~\bibnamefont{Derevianko}} \bibnamefont{and}
  \bibinfo{author}{\bibfnamefont{C.~C.} \bibnamefont{Cannon}},
  \bibinfo{journal}{Phys. Rev. A} \textbf{\bibinfo{volume}{70}},
  \bibinfo{pages}{062319} (\bibinfo{year}{2004}).

\bibitem[{\citenamefont{Rabl et~al.}(2006)\citenamefont{Rabl, DeMille, Doyle,
  Lukin, Schoelkopf, and Zoller}}]{PRable:06}
\bibinfo{author}{\bibfnamefont{P.}~\bibnamefont{Rabl}},
  \bibinfo{author}{\bibfnamefont{D.}~\bibnamefont{DeMille}},
  \bibinfo{author}{\bibfnamefont{J.~M.} \bibnamefont{Doyle}},
  \bibinfo{author}{\bibfnamefont{M.~D.} \bibnamefont{Lukin}},
  \bibinfo{author}{\bibfnamefont{R.~J.} \bibnamefont{Schoelkopf}},
  \bibnamefont{and} \bibinfo{author}{\bibfnamefont{P.}~\bibnamefont{Zoller}},
  \bibinfo{journal}{Phys. Rev. Lett.} \textbf{\bibinfo{volume}{97}},
  \bibinfo{pages}{033003} (\bibinfo{year}{2006}).

\bibitem[{\citenamefont{Hudson et~al.}(2002)\citenamefont{Hudson, Sauer,
  Tarbutt, and Hinds}}]{JJHudson:02}
\bibinfo{author}{\bibfnamefont{J.~J.} \bibnamefont{Hudson}},
  \bibinfo{author}{\bibfnamefont{B.~E.} \bibnamefont{Sauer}},
  \bibinfo{author}{\bibfnamefont{M.~R.} \bibnamefont{Tarbutt}},
  \bibnamefont{and} \bibinfo{author}{\bibfnamefont{E.~A.} \bibnamefont{Hinds}},
  \bibinfo{journal}{Phys. Rev. Lett.} \textbf{\bibinfo{volume}{89}},
  \bibinfo{pages}{023003} (\bibinfo{year}{2002}).

\bibitem[{\citenamefont{Zelevinsky et~al.}(2008)\citenamefont{Zelevinsky,
  Kotochigova, and Ye}}]{TZelevinsky:08}
\bibinfo{author}{\bibfnamefont{T.}~\bibnamefont{Zelevinsky}},
  \bibinfo{author}{\bibfnamefont{S.}~\bibnamefont{Kotochigova}},
  \bibnamefont{and} \bibinfo{author}{\bibfnamefont{J.}~\bibnamefont{Ye}},
  \bibinfo{journal}{Phys. Rev. Lett.} \textbf{\bibinfo{volume}{100}},
  \bibinfo{pages}{043201} (\bibinfo{year}{2008}).

\bibitem[{\citenamefont{Bloom et~al.}(2014)\citenamefont{Bloom, Nicholson,
  Williams, Campbell, Bishof, Zhang, Zhang, Bromley, and Ye}}]{BJBloom:14}
\bibinfo{author}{\bibfnamefont{B.~J.} \bibnamefont{Bloom}},
  \bibinfo{author}{\bibfnamefont{T.~L.} \bibnamefont{Nicholson}},
  \bibinfo{author}{\bibfnamefont{J.~R.} \bibnamefont{Williams}},
  \bibinfo{author}{\bibfnamefont{S.~L.} \bibnamefont{Campbell}},
  \bibinfo{author}{\bibfnamefont{M.}~\bibnamefont{Bishof}},
  \bibinfo{author}{\bibfnamefont{X.}~\bibnamefont{Zhang}},
  \bibinfo{author}{\bibfnamefont{W.}~\bibnamefont{Zhang}},
  \bibinfo{author}{\bibfnamefont{S.~L.} \bibnamefont{Bromley}},
  \bibnamefont{and} \bibinfo{author}{\bibfnamefont{J.}~\bibnamefont{Ye}},
  \bibinfo{journal}{Nature} \textbf{\bibinfo{volume}{506}}, \bibinfo{pages}{71}
  (\bibinfo{year}{2014}).

\bibitem[{\citenamefont{Chin et~al.}(2010)\citenamefont{Chin, Grimm, Julienne,
  and Tiesinga}}]{CChin:10}
\bibinfo{author}{\bibfnamefont{C.}~\bibnamefont{Chin}},
  \bibinfo{author}{\bibfnamefont{R.}~\bibnamefont{Grimm}},
  \bibinfo{author}{\bibfnamefont{P.}~\bibnamefont{Julienne}}, \bibnamefont{and}
  \bibinfo{author}{\bibfnamefont{E.}~\bibnamefont{Tiesinga}},
  \bibinfo{journal}{Rev. Mod. Phys.} \textbf{\bibinfo{volume}{82}},
  \bibinfo{pages}{1225} (\bibinfo{year}{2010}).

\bibitem[{\citenamefont{Hutson and Sold\'an}(2006)}]{JMHutson:06}
\bibinfo{author}{\bibfnamefont{J.~M.} \bibnamefont{Hutson}} \bibnamefont{and}
  \bibinfo{author}{\bibfnamefont{P.}~\bibnamefont{Sold\'an}},
  \bibinfo{journal}{Int. Rev. Phys. Chem.} \textbf{\bibinfo{volume}{25}},
  \bibinfo{pages}{497} (\bibinfo{year}{2006}).

\bibitem[{\citenamefont{K{\"o}hler et~al.}(2006)\citenamefont{K{\"o}hler,
  G\'oral, and Julienne}}]{TKohler:06}
\bibinfo{author}{\bibfnamefont{T.}~\bibnamefont{K{\"o}hler}},
  \bibinfo{author}{\bibfnamefont{K.}~\bibnamefont{G\'oral}}, \bibnamefont{and}
  \bibinfo{author}{\bibfnamefont{P.~S.} \bibnamefont{Julienne}},
  \bibinfo{journal}{Rev. Mod. Phys.} \textbf{\bibinfo{volume}{78}},
  \bibinfo{pages}{1311} (\bibinfo{year}{2006}).

\bibitem[{\citenamefont{Jones et~al.}(2006)\citenamefont{Jones, Tiesinga, Lett,
  and Julienne}}]{KMJones:06}
\bibinfo{author}{\bibfnamefont{K.~M.} \bibnamefont{Jones}},
  \bibinfo{author}{\bibfnamefont{E.}~\bibnamefont{Tiesinga}},
  \bibinfo{author}{\bibfnamefont{P.~D.} \bibnamefont{Lett}}, \bibnamefont{and}
  \bibinfo{author}{\bibfnamefont{P.~S.} \bibnamefont{Julienne}},
  \bibinfo{journal}{Rev. Mod. Phys.} \textbf{\bibinfo{volume}{78}},
  \bibinfo{pages}{483} (\bibinfo{year}{2006}).

\bibitem[{\citenamefont{Lang et~al.}(2008)\citenamefont{Lang, Winkler, Strauss,
  Grimm, and Denschlag}}]{FLang:08}
\bibinfo{author}{\bibfnamefont{F.}~\bibnamefont{Lang}},
  \bibinfo{author}{\bibfnamefont{K.}~\bibnamefont{Winkler}},
  \bibinfo{author}{\bibfnamefont{C.}~\bibnamefont{Strauss}},
  \bibinfo{author}{\bibfnamefont{R.}~\bibnamefont{Grimm}}, \bibnamefont{and}
  \bibinfo{author}{\bibfnamefont{J.~H.} \bibnamefont{Denschlag}},
  \bibinfo{journal}{Phys. Rev. Lett.} \textbf{\bibinfo{volume}{101}},
  \bibinfo{pages}{133005} (\bibinfo{year}{2008}).

\bibitem[{\citenamefont{Ni et~al.}(2008)\citenamefont{Ni, Ospelkaus, {de
  Miranda}, Pe'er, Neyenhuis, Zirbel, Kotochigova, Julienne, Jin, and
  Ye}}]{K-KNi:08}
\bibinfo{author}{\bibfnamefont{K.-K.} \bibnamefont{Ni}},
  \bibinfo{author}{\bibfnamefont{S.}~\bibnamefont{Ospelkaus}},
  \bibinfo{author}{\bibfnamefont{M.~H.~G.} \bibnamefont{{de Miranda}}},
  \bibinfo{author}{\bibfnamefont{A.}~\bibnamefont{Pe'er}},
  \bibinfo{author}{\bibfnamefont{B.}~\bibnamefont{Neyenhuis}},
  \bibinfo{author}{\bibfnamefont{J.~J.} \bibnamefont{Zirbel}},
  \bibinfo{author}{\bibfnamefont{S.}~\bibnamefont{Kotochigova}},
  \bibinfo{author}{\bibfnamefont{P.~S.} \bibnamefont{Julienne}},
  \bibinfo{author}{\bibfnamefont{D.~S.} \bibnamefont{Jin}}, \bibnamefont{and}
  \bibinfo{author}{\bibfnamefont{J.}~\bibnamefont{Ye}},
  \bibinfo{journal}{Science} \textbf{\bibinfo{volume}{322}},
  \bibinfo{pages}{231} (\bibinfo{year}{2008}).

\bibitem[{\citenamefont{Danzl et~al.}(2010)\citenamefont{Danzl, Mark, Haller,
  Gustavsson, Hart, Aldegunde, Hutson, and N{\"a}gerl}}]{JGDanzl:10}
\bibinfo{author}{\bibfnamefont{J.~G.} \bibnamefont{Danzl}},
  \bibinfo{author}{\bibfnamefont{M.~J.} \bibnamefont{Mark}},
  \bibinfo{author}{\bibfnamefont{E.}~\bibnamefont{Haller}},
  \bibinfo{author}{\bibfnamefont{M.}~\bibnamefont{Gustavsson}},
  \bibinfo{author}{\bibfnamefont{R.}~\bibnamefont{Hart}},
  \bibinfo{author}{\bibfnamefont{J.}~\bibnamefont{Aldegunde}},
  \bibinfo{author}{\bibfnamefont{J.~M.} \bibnamefont{Hutson}},
  \bibnamefont{and} \bibinfo{author}{\bibfnamefont{H.-C.}
  \bibnamefont{N{\"a}gerl}}, \bibinfo{journal}{Nat. Phys.}
  \textbf{\bibinfo{volume}{6}}, \bibinfo{pages}{265} (\bibinfo{year}{2010}).

\bibitem[{\citenamefont{Takekoshi et~al.}(2014)\citenamefont{Takekoshi,
  Reichs\"ollner, Schindewolf, Hutson, Le~Sueur, Dulieu, Ferlaino, Grimm, and
  N\"agerl}}]{TTakekoshi:14a}
\bibinfo{author}{\bibfnamefont{T.}~\bibnamefont{Takekoshi}},
  \bibinfo{author}{\bibfnamefont{L.}~\bibnamefont{Reichs\"ollner}},
  \bibinfo{author}{\bibfnamefont{A.}~\bibnamefont{Schindewolf}},
  \bibinfo{author}{\bibfnamefont{J.~M.} \bibnamefont{Hutson}},
  \bibinfo{author}{\bibfnamefont{C.~R.} \bibnamefont{Le~Sueur}},
  \bibinfo{author}{\bibfnamefont{O.}~\bibnamefont{Dulieu}},
  \bibinfo{author}{\bibfnamefont{F.}~\bibnamefont{Ferlaino}},
  \bibinfo{author}{\bibfnamefont{R.}~\bibnamefont{Grimm}}, \bibnamefont{and}
  \bibinfo{author}{\bibfnamefont{H.-C.} \bibnamefont{N\"agerl}},
  \bibinfo{journal}{Phys. Rev. Lett.} \textbf{\bibinfo{volume}{113}},
  \bibinfo{pages}{205301} (\bibinfo{year}{2014}).

\bibitem[{\citenamefont{Griesmaier et~al.}(2005)\citenamefont{Griesmaier,
  Werner, Hensler, Stuhler, and Pfau}}]{AGriesmaier:05}
\bibinfo{author}{\bibfnamefont{A.}~\bibnamefont{Griesmaier}},
  \bibinfo{author}{\bibfnamefont{J.}~\bibnamefont{Werner}},
  \bibinfo{author}{\bibfnamefont{S.}~\bibnamefont{Hensler}},
  \bibinfo{author}{\bibfnamefont{J.}~\bibnamefont{Stuhler}}, \bibnamefont{and}
  \bibinfo{author}{\bibfnamefont{T.}~\bibnamefont{Pfau}},
  \bibinfo{journal}{Phys. Rev. Lett.} \textbf{\bibinfo{volume}{94}},
  \bibinfo{pages}{160401} (\bibinfo{year}{2005}).

\bibitem[{\citenamefont{de~Paz et~al.}(2013)\citenamefont{de~Paz, Sharma,
  Chotia, Mar\'echal, Huckans, Pedri, Santos, Gorceix, Vernac, and
  Laburthe-Tolra}}]{ADePaz:13}
\bibinfo{author}{\bibfnamefont{A.}~\bibnamefont{de~Paz}},
  \bibinfo{author}{\bibfnamefont{A.}~\bibnamefont{Sharma}},
  \bibinfo{author}{\bibfnamefont{A.}~\bibnamefont{Chotia}},
  \bibinfo{author}{\bibfnamefont{E.}~\bibnamefont{Mar\'echal}},
  \bibinfo{author}{\bibfnamefont{J.~H.} \bibnamefont{Huckans}},
  \bibinfo{author}{\bibfnamefont{P.}~\bibnamefont{Pedri}},
  \bibinfo{author}{\bibfnamefont{L.}~\bibnamefont{Santos}},
  \bibinfo{author}{\bibfnamefont{O.}~\bibnamefont{Gorceix}},
  \bibinfo{author}{\bibfnamefont{L.}~\bibnamefont{Vernac}}, \bibnamefont{and}
  \bibinfo{author}{\bibfnamefont{B.}~\bibnamefont{Laburthe-Tolra}},
  \bibinfo{journal}{Phys. Rev. Lett.} \textbf{\bibinfo{volume}{111}},
  \bibinfo{pages}{185305} (\bibinfo{year}{2013}).

\bibitem[{\citenamefont{Lu et~al.}(2011)\citenamefont{Lu, Burdick, Youn, and
  Lev}}]{MLu:11}
\bibinfo{author}{\bibfnamefont{M.}~\bibnamefont{Lu}},
  \bibinfo{author}{\bibfnamefont{N.~Q.} \bibnamefont{Burdick}},
  \bibinfo{author}{\bibfnamefont{S.~H.} \bibnamefont{Youn}}, \bibnamefont{and}
  \bibinfo{author}{\bibfnamefont{B.~L.} \bibnamefont{Lev}},
  \bibinfo{journal}{Phys. Rev. Lett.} \textbf{\bibinfo{volume}{107}},
  \bibinfo{pages}{190401} (\bibinfo{year}{2011}).

\bibitem[{\citenamefont{Lu et~al.}(2012)\citenamefont{Lu, Burdick, and
  Lev}}]{MLu:12}
\bibinfo{author}{\bibfnamefont{M.}~\bibnamefont{Lu}},
  \bibinfo{author}{\bibfnamefont{N.~Q.} \bibnamefont{Burdick}},
  \bibnamefont{and} \bibinfo{author}{\bibfnamefont{B.~L.} \bibnamefont{Lev}},
  \bibinfo{journal}{Phys. Rev. Lett.} \textbf{\bibinfo{volume}{108}},
  \bibinfo{pages}{215301} (\bibinfo{year}{2012}).

\bibitem[{\citenamefont{Aikawa et~al.}(2012)\citenamefont{Aikawa, Frisch, Mark,
  Baier, Rietzler, Grimm, and Ferlaino}}]{KAikawa:12}
\bibinfo{author}{\bibfnamefont{K.}~\bibnamefont{Aikawa}},
  \bibinfo{author}{\bibfnamefont{A.}~\bibnamefont{Frisch}},
  \bibinfo{author}{\bibfnamefont{M.}~\bibnamefont{Mark}},
  \bibinfo{author}{\bibfnamefont{S.}~\bibnamefont{Baier}},
  \bibinfo{author}{\bibfnamefont{A.}~\bibnamefont{Rietzler}},
  \bibinfo{author}{\bibfnamefont{R.}~\bibnamefont{Grimm}}, \bibnamefont{and}
  \bibinfo{author}{\bibfnamefont{F.}~\bibnamefont{Ferlaino}},
  \bibinfo{journal}{Phys. Rev. Lett.} \textbf{\bibinfo{volume}{108}},
  \bibinfo{pages}{210401} (\bibinfo{year}{2012}).

\bibitem[{\citenamefont{Aikawa et~al.}(2014{\natexlab{a}})\citenamefont{Aikawa,
  Frisch, Mark, Baier, Grimm, and Ferlaino}}]{KAikawa:14a}
\bibinfo{author}{\bibfnamefont{K.}~\bibnamefont{Aikawa}},
  \bibinfo{author}{\bibfnamefont{A.}~\bibnamefont{Frisch}},
  \bibinfo{author}{\bibfnamefont{M.}~\bibnamefont{Mark}},
  \bibinfo{author}{\bibfnamefont{S.}~\bibnamefont{Baier}},
  \bibinfo{author}{\bibfnamefont{R.}~\bibnamefont{Grimm}}, \bibnamefont{and}
  \bibinfo{author}{\bibfnamefont{F.}~\bibnamefont{Ferlaino}},
  \bibinfo{journal}{Phys. Rev. Lett.} \textbf{\bibinfo{volume}{112}},
  \bibinfo{pages}{010404} (\bibinfo{year}{2014}{\natexlab{a}}).

\bibitem[{\citenamefont{Kotochigova}(2014)}]{SKotochigova:14a}
\bibinfo{author}{\bibfnamefont{S.}~\bibnamefont{Kotochigova}},
  \bibinfo{journal}{Rep. Prog. Phys.} \textbf{\bibinfo{volume}{77}},
  \bibinfo{pages}{093901} (\bibinfo{year}{2014}).

\bibitem[{\citenamefont{Petrov et~al.}(2012)\citenamefont{Petrov, Tiesinga, and
  Kotochigova}}]{APetrov:12}
\bibinfo{author}{\bibfnamefont{A.}~\bibnamefont{Petrov}},
  \bibinfo{author}{\bibfnamefont{E.}~\bibnamefont{Tiesinga}}, \bibnamefont{and}
  \bibinfo{author}{\bibfnamefont{S.}~\bibnamefont{Kotochigova}},
  \bibinfo{journal}{Phys. Rev. Lett.} \textbf{\bibinfo{volume}{109}},
  \bibinfo{pages}{103002} (\bibinfo{year}{2012}).

\bibitem[{\citenamefont{Frisch et~al.}(2014)\citenamefont{Frisch, Mark, Aikawa,
  Ferlaino, Bohn, Makrides, Petrov, and Kotochigova}}]{AFrisch:14}
\bibinfo{author}{\bibfnamefont{A.}~\bibnamefont{Frisch}},
  \bibinfo{author}{\bibfnamefont{M.}~\bibnamefont{Mark}},
  \bibinfo{author}{\bibfnamefont{K.}~\bibnamefont{Aikawa}},
  \bibinfo{author}{\bibfnamefont{F.}~\bibnamefont{Ferlaino}},
  \bibinfo{author}{\bibfnamefont{J.~L.} \bibnamefont{Bohn}},
  \bibinfo{author}{\bibfnamefont{C.}~\bibnamefont{Makrides}},
  \bibinfo{author}{\bibfnamefont{A.}~\bibnamefont{Petrov}}, \bibnamefont{and}
  \bibinfo{author}{\bibfnamefont{S.}~\bibnamefont{Kotochigova}},
  \bibinfo{journal}{Nature} \textbf{\bibinfo{volume}{507}},
  \bibinfo{pages}{475} (\bibinfo{year}{2014}).

\bibitem[{\citenamefont{Baumann et~al.}(2014)\citenamefont{Baumann, Burdick,
  Lu, and Lev}}]{KBaumann:14}
\bibinfo{author}{\bibfnamefont{K.}~\bibnamefont{Baumann}},
  \bibinfo{author}{\bibfnamefont{N.~Q.} \bibnamefont{Burdick}},
  \bibinfo{author}{\bibfnamefont{M.}~\bibnamefont{Lu}}, \bibnamefont{and}
  \bibinfo{author}{\bibfnamefont{B.~L.} \bibnamefont{Lev}},
  \bibinfo{journal}{Phys. Rev. A} \textbf{\bibinfo{volume}{89}},
  \bibinfo{pages}{020701(R)} (\bibinfo{year}{2014}).

\bibitem[{\citenamefont{Qu\'em\'ener and Julienne}(2012)}]{GQuemener:12}
\bibinfo{author}{\bibfnamefont{G.}~\bibnamefont{Qu\'em\'ener}}
  \bibnamefont{and} \bibinfo{author}{\bibfnamefont{P.~S.}
  \bibnamefont{Julienne}}, \bibinfo{journal}{Chem. Rev.}
  \textbf{\bibinfo{volume}{112}}, \bibinfo{pages}{4949} (\bibinfo{year}{2012}).

\bibitem[{\citenamefont{Yan et~al.}(2013)\citenamefont{Yan, Moses, Gadway,
  Covey, Hazzard, Rey, Jin, and Ye}}]{BYan:13}
\bibinfo{author}{\bibfnamefont{B.}~\bibnamefont{Yan}},
  \bibinfo{author}{\bibfnamefont{S.~A.} \bibnamefont{Moses}},
  \bibinfo{author}{\bibfnamefont{B.}~\bibnamefont{Gadway}},
  \bibinfo{author}{\bibfnamefont{J.~P.} \bibnamefont{Covey}},
  \bibinfo{author}{\bibfnamefont{K.~R.~A.} \bibnamefont{Hazzard}},
  \bibinfo{author}{\bibfnamefont{A.~M.} \bibnamefont{Rey}},
  \bibinfo{author}{\bibfnamefont{D.~S.} \bibnamefont{Jin}}, \bibnamefont{and}
  \bibinfo{author}{\bibfnamefont{J.}~\bibnamefont{Ye}},
  \bibinfo{journal}{Nature} \textbf{\bibinfo{volume}{501}},
  \bibinfo{pages}{521} (\bibinfo{year}{2013}).

\bibitem[{\citenamefont{Aikawa et~al.}(2014{\natexlab{b}})\citenamefont{Aikawa,
  Frisch, Mark, Baier, Grimm, Bohn, Jin, Bruun, and Ferlaino}}]{KAikawa:14b}
\bibinfo{author}{\bibfnamefont{K.}~\bibnamefont{Aikawa}},
  \bibinfo{author}{\bibfnamefont{A.}~\bibnamefont{Frisch}},
  \bibinfo{author}{\bibfnamefont{M.}~\bibnamefont{Mark}},
  \bibinfo{author}{\bibfnamefont{S.}~\bibnamefont{Baier}},
  \bibinfo{author}{\bibfnamefont{R.}~\bibnamefont{Grimm}},
  \bibinfo{author}{\bibfnamefont{J.~L.} \bibnamefont{Bohn}},
  \bibinfo{author}{\bibfnamefont{D.~S.} \bibnamefont{Jin}},
  \bibinfo{author}{\bibfnamefont{G.~M.} \bibnamefont{Bruun}}, \bibnamefont{and}
  \bibinfo{author}{\bibfnamefont{F.}~\bibnamefont{Ferlaino}},
  \bibinfo{journal}{Phys. Rev. Lett.} \textbf{\bibinfo{volume}{113}},
  \bibinfo{pages}{263201} (\bibinfo{year}{2014}{\natexlab{b}}).

\bibitem[{\citenamefont{Gonz\'alez-Mart\'{\i}nez and
  Hutson}(2013{\natexlab{a}})}]{MLGonzalez-Martinez:13a}
\bibinfo{author}{\bibfnamefont{M.~L.} \bibnamefont{Gonz\'alez-Mart\'{\i}nez}}
  \bibnamefont{and} \bibinfo{author}{\bibfnamefont{J.~M.}
  \bibnamefont{Hutson}}, \bibinfo{journal}{Phys. Rev. A}
  \textbf{\bibinfo{volume}{88}}, \bibinfo{pages}{020701(R)}
  (\bibinfo{year}{2013}{\natexlab{a}}).

\bibitem[{Note1()}]{Note1}
\bibinfo{note}{Preliminary multireference configuration interaction
  (MRCI) calculations predict a permanent electric dipole moment of about
  1.5~Debye for the lowest $^4\Sigma $ state, and the magnetic dipole moment
  may be up to approximately 8 Bohr magnetons.}

\bibitem[{\citenamefont{Tomza}(2013)}]{MTomza:13b}
\bibinfo{author}{\bibfnamefont{M.}~\bibnamefont{Tomza}},
  \bibinfo{journal}{Phys. Rev. A} \textbf{\bibinfo{volume}{88}},
  \bibinfo{pages}{012519} (\bibinfo{year}{2013}).

\bibitem[{\citenamefont{Tomza}(2014)}]{MTomza:14a}
\bibinfo{author}{\bibfnamefont{M.}~\bibnamefont{Tomza}},
  \bibinfo{journal}{Phys. Rev. A} \textbf{\bibinfo{volume}{90}},
  \bibinfo{pages}{022514} (\bibinfo{year}{2014}).

\bibitem[{\citenamefont{Pires et~al.}(2014)\citenamefont{Pires, Ulmanis,
  H\"afner, Repp, Arias, Kuhnle, and Weidem\"uller}}]{RPires:14a}
\bibinfo{author}{\bibfnamefont{R.}~\bibnamefont{Pires}},
  \bibinfo{author}{\bibfnamefont{J.}~\bibnamefont{Ulmanis}},
  \bibinfo{author}{\bibfnamefont{S.}~\bibnamefont{H\"afner}},
  \bibinfo{author}{\bibfnamefont{M.}~\bibnamefont{Repp}},
  \bibinfo{author}{\bibfnamefont{A.}~\bibnamefont{Arias}},
  \bibinfo{author}{\bibfnamefont{E.~D.} \bibnamefont{Kuhnle}},
  \bibnamefont{and}
  \bibinfo{author}{\bibfnamefont{M.}~\bibnamefont{Weidem\"uller}},
  \bibinfo{journal}{Phys. Rev. Lett.} \textbf{\bibinfo{volume}{112}},
  \bibinfo{pages}{250404} (\bibinfo{year}{2014}).

\bibitem[{\citenamefont{Gonz\'alez-Mart\'{\i}nez and
  Hutson}(2013{\natexlab{b}})}]{MLGonzalez-Martinez:13c}
\bibinfo{author}{\bibfnamefont{M.~L.} \bibnamefont{Gonz\'alez-Mart\'{\i}nez}}
  \bibnamefont{and} \bibinfo{author}{\bibfnamefont{J.~M.}
  \bibnamefont{Hutson}}, \bibinfo{journal}{Phys. Rev. A}
  \textbf{\bibinfo{volume}{88}}, \bibinfo{pages}{053420}
  (\bibinfo{year}{2013}{\natexlab{b}}).

\bibitem[{\citenamefont{Beckmann et~al.}(1974)\citenamefont{Beckmann, B\"oklen,
  and Elke}}]{ABeckmann:74}
\bibinfo{author}{\bibfnamefont{A.}~\bibnamefont{Beckmann}},
  \bibinfo{author}{\bibfnamefont{K.~D.} \bibnamefont{B\"oklen}},
  \bibnamefont{and} \bibinfo{author}{\bibfnamefont{D.}~\bibnamefont{Elke}},
  \bibinfo{journal}{Z. Phys.} \textbf{\bibinfo{volume}{270}},
  \bibinfo{pages}{173} (\bibinfo{year}{1974}).

\bibitem[{\citenamefont{Stone}(2005)}]{NJStone:05}
\bibinfo{author}{\bibfnamefont{N.~J.} \bibnamefont{Stone}},
  \bibinfo{journal}{At. Data Nucl. Data Tables} \textbf{\bibinfo{volume}{90}},
  \bibinfo{pages}{75} (\bibinfo{year}{2005}).

\bibitem[{\citenamefont{Martin et~al.}(1978)\citenamefont{Martin, Zalubas, and
  Hagan}}]{WCMartin:78}
\bibinfo{author}{\bibfnamefont{W.~C.} \bibnamefont{Martin}},
  \bibinfo{author}{\bibfnamefont{R.}~\bibnamefont{Zalubas}}, \bibnamefont{and}
  \bibinfo{author}{\bibfnamefont{L.}~\bibnamefont{Hagan}},
  \bibinfo{journal}{Natl. Stand. Ref. Data Ser., Natl. Bur. Stand. (U.S.)}
  \textbf{\bibinfo{volume}{60}} (\bibinfo{year}{1978}).

\bibitem[{\citenamefont{Krems et~al.}(2004)\citenamefont{Krems, Groenenboom,
  and Dalgarno}}]{RVKrems:04b}
\bibinfo{author}{\bibfnamefont{R.~V.} \bibnamefont{Krems}},
  \bibinfo{author}{\bibfnamefont{G.~C.} \bibnamefont{Groenenboom}},
  \bibnamefont{and} \bibinfo{author}{\bibfnamefont{A.}~\bibnamefont{Dalgarno}},
  \bibinfo{journal}{J. Phys. Chem. A} \textbf{\bibinfo{volume}{108}},
  \bibinfo{pages}{8941} (\bibinfo{year}{2004}).

\bibitem[{\citenamefont{Werner et~al.}(2012)\citenamefont{Werner, Knowles,
  Knizia, Manby, {Sch\"{u}tz} et~al.}}]{MOLPRO2012_brief}
\bibinfo{author}{\bibfnamefont{H.-J.} \bibnamefont{Werner}},
  \bibinfo{author}{\bibfnamefont{P.~J.} \bibnamefont{Knowles}},
  \bibinfo{author}{\bibfnamefont{G.}~\bibnamefont{Knizia}},
  \bibinfo{author}{\bibfnamefont{F.~R.} \bibnamefont{Manby}},
  \bibinfo{author}{\bibfnamefont{M.}~\bibnamefont{{Sch\"{u}tz}}},
  \bibnamefont{et~al.}, \emph{\bibinfo{title}{\textsc{MOLPRO}, version 2012.1:
  A package of ab initio programs}} (\bibinfo{year}{2012}), \bibinfo{note}{see
  http://www.molpro.net}.

\bibitem[{\citenamefont{Prascher et~al.}(2011)\citenamefont{Prascher, Woon,
  Peterson, Dunning~Jr., and Wilson}}]{BPPrascher:11}
\bibinfo{author}{\bibfnamefont{B.~P.} \bibnamefont{Prascher}},
  \bibinfo{author}{\bibfnamefont{D.~E.} \bibnamefont{Woon}},
  \bibinfo{author}{\bibfnamefont{K.~A.} \bibnamefont{Peterson}},
  \bibinfo{author}{\bibfnamefont{T.~H.} \bibnamefont{Dunning~Jr.}},
  \bibnamefont{and} \bibinfo{author}{\bibfnamefont{A.~K.}
  \bibnamefont{Wilson}}, \bibinfo{journal}{Theor. Chem. Acc.}
  \textbf{\bibinfo{volume}{128}}, \bibinfo{pages}{69} (\bibinfo{year}{2011}).

\bibitem[{\citenamefont{Dolg et~al.}(1989)\citenamefont{Dolg, Stoll, and
  Preuss}}]{MDolg:89}
\bibinfo{author}{\bibfnamefont{M.}~\bibnamefont{Dolg}},
  \bibinfo{author}{\bibfnamefont{H.}~\bibnamefont{Stoll}}, \bibnamefont{and}
  \bibinfo{author}{\bibfnamefont{H.}~\bibnamefont{Preuss}},
  \bibinfo{journal}{J. Chem. Phys.} \textbf{\bibinfo{volume}{90}},
  \bibinfo{pages}{1730} (\bibinfo{year}{1989}).

\bibitem[{\citenamefont{Jeziorski et~al.}(1994)\citenamefont{Jeziorski,
  Moszynski, and Szalewicz}}]{BJeziorski:94}
\bibinfo{author}{\bibfnamefont{B.}~\bibnamefont{Jeziorski}},
  \bibinfo{author}{\bibfnamefont{R.}~\bibnamefont{Moszynski}},
  \bibnamefont{and}
  \bibinfo{author}{\bibfnamefont{K.}~\bibnamefont{Szalewicz}},
  \bibinfo{journal}{Chem. Rev.} \textbf{\bibinfo{volume}{94}},
  \bibinfo{pages}{1887} (\bibinfo{year}{1994}).

\bibitem[{\citenamefont{Zhang et~al.}(2010)\citenamefont{Zhang, Sadeghpour, and
  Dalgarno}}]{PZhang:10}
\bibinfo{author}{\bibfnamefont{P.}~\bibnamefont{Zhang}},
  \bibinfo{author}{\bibfnamefont{H.~R.} \bibnamefont{Sadeghpour}},
  \bibnamefont{and} \bibinfo{author}{\bibfnamefont{A.}~\bibnamefont{Dalgarno}},
  \bibinfo{journal}{J. Chem. Phys.} \textbf{\bibinfo{volume}{133}},
  \bibinfo{pages}{044306} (\bibinfo{year}{2010}).

\bibitem[{\citenamefont{Heisenberg}(1928)}]{WHeisenberg:28}
\bibinfo{author}{\bibfnamefont{W.}~\bibnamefont{Heisenberg}},
  \bibinfo{journal}{Z. Phys.} \textbf{\bibinfo{volume}{49}},
  \bibinfo{pages}{619} (\bibinfo{year}{1928}).

\bibitem[{\citenamefont{Buchachenko et~al.}(2009)\citenamefont{Buchachenko,
  Cha\l{}asi\'nski, and Szcz\c{e}\'sniak}}]{AABuchachenko:09}
\bibinfo{author}{\bibfnamefont{A.~A.} \bibnamefont{Buchachenko}},
  \bibinfo{author}{\bibfnamefont{G.}~\bibnamefont{Cha\l{}asi\'nski}},
  \bibnamefont{and} \bibinfo{author}{\bibfnamefont{M.~M.}
  \bibnamefont{Szcz\c{e}\'sniak}}, \bibinfo{journal}{J. Chem. Phys.}
  \textbf{\bibinfo{volume}{131}}, \bibinfo{pages}{241102}
  (\bibinfo{year}{2009}).

\bibitem[{\citenamefont{Tscherbul et~al.}(2010)\citenamefont{Tscherbul,
  K\l{}os, Dalgarno, Zygelman, Pavlovic, Hummon, Lu, Tsikata, and
  Doyle}}]{TVTscherbul:10a}
\bibinfo{author}{\bibfnamefont{T.~V.} \bibnamefont{Tscherbul}},
  \bibinfo{author}{\bibfnamefont{J.}~\bibnamefont{K\l{}os}},
  \bibinfo{author}{\bibfnamefont{A.}~\bibnamefont{Dalgarno}},
  \bibinfo{author}{\bibfnamefont{B.}~\bibnamefont{Zygelman}},
  \bibinfo{author}{\bibfnamefont{Z.}~\bibnamefont{Pavlovic}},
  \bibinfo{author}{\bibfnamefont{M.~T.} \bibnamefont{Hummon}},
  \bibinfo{author}{\bibfnamefont{H.-I.} \bibnamefont{Lu}},
  \bibinfo{author}{\bibfnamefont{E.}~\bibnamefont{Tsikata}}, \bibnamefont{and}
  \bibinfo{author}{\bibfnamefont{J.~M.} \bibnamefont{Doyle}},
  \bibinfo{journal}{Phys. Rev. A} \textbf{\bibinfo{volume}{82}},
  \bibinfo{pages}{042718} (\bibinfo{year}{2010}).

\bibitem[{\citenamefont{Ho and Rabitz}(1996)}]{T-SHo:96a}
\bibinfo{author}{\bibfnamefont{T.-S.} \bibnamefont{Ho}} \bibnamefont{and}
  \bibinfo{author}{\bibfnamefont{H.}~\bibnamefont{Rabitz}},
  \bibinfo{journal}{J. Chem. Phys.} \textbf{\bibinfo{volume}{104}},
  \bibinfo{pages}{2584} (\bibinfo{year}{1996}).

\bibitem[{\citenamefont{Tang}(1969)}]{KTTang:69}
\bibinfo{author}{\bibfnamefont{K.~T.} \bibnamefont{Tang}},
  \bibinfo{journal}{Phys. Rev.} \textbf{\bibinfo{volume}{177}},
  \bibinfo{pages}{108} (\bibinfo{year}{1969}).

\bibitem[{\citenamefont{Derevianko et~al.}(2010)\citenamefont{Derevianko,
  Porsev, and Babb}}]{ADerevianko:10}
\bibinfo{author}{\bibfnamefont{A.}~\bibnamefont{Derevianko}},
  \bibinfo{author}{\bibfnamefont{S.~G.} \bibnamefont{Porsev}},
  \bibnamefont{and} \bibinfo{author}{\bibfnamefont{J.~F.} \bibnamefont{Babb}},
  \bibinfo{journal}{At. Data Nucl. Data Tables} \textbf{\bibinfo{volume}{96}},
  \bibinfo{pages}{323} (\bibinfo{year}{2010}).

\bibitem[{\citenamefont{Lepers et~al.}(2014)\citenamefont{Lepers, Wyart, and
  Dulieu}}]{MLepers:14a}
\bibinfo{author}{\bibfnamefont{M.}~\bibnamefont{Lepers}},
  \bibinfo{author}{\bibfnamefont{J.-F.} \bibnamefont{Wyart}}, \bibnamefont{and}
  \bibinfo{author}{\bibfnamefont{O.}~\bibnamefont{Dulieu}},
  \bibinfo{journal}{Phys. Rev. A} \textbf{\bibinfo{volume}{89}},
  \bibinfo{pages}{022505} (\bibinfo{year}{2014}).

\bibitem[{\citenamefont{Chu et~al.}(2007)\citenamefont{Chu, Dalgarno, and
  Groenenboom}}]{XChu:07}
\bibinfo{author}{\bibfnamefont{X.}~\bibnamefont{Chu}},
  \bibinfo{author}{\bibfnamefont{A.}~\bibnamefont{Dalgarno}}, \bibnamefont{and}
  \bibinfo{author}{\bibfnamefont{G.~C.} \bibnamefont{Groenenboom}},
  \bibinfo{journal}{Phys. Rev. A} \textbf{\bibinfo{volume}{75}},
  \bibinfo{pages}{032723} (\bibinfo{year}{2007}).

\bibitem[{\citenamefont{Krems et~al.}(2005)\citenamefont{Krems, K\l{}os, Rode,
  Szcz\ifmmode \mbox{\c{e}}\else \c{e}\fi{}\ifmmode~\acute{s}\else
  \'{s}\fi{}niak, Cha\l{}asi\ifmmode~\acute{n}\else \'{n}\fi{}ski, and
  Dalgarno}}]{RVKrems:05b}
\bibinfo{author}{\bibfnamefont{R.~V.} \bibnamefont{Krems}},
  \bibinfo{author}{\bibfnamefont{J.}~\bibnamefont{K\l{}os}},
  \bibinfo{author}{\bibfnamefont{M.~F.} \bibnamefont{Rode}},
  \bibinfo{author}{\bibfnamefont{M.~M.} \bibnamefont{Szcz\ifmmode
  \mbox{\c{e}}\else \c{e}\fi{}\ifmmode~\acute{s}\else \'{s}\fi{}niak}},
  \bibinfo{author}{\bibfnamefont{G.}~\bibnamefont{Cha\l{}asi\ifmmode~\acute{n}\else
  \'{n}\fi{}ski}}, \bibnamefont{and}
  \bibinfo{author}{\bibfnamefont{A.}~\bibnamefont{Dalgarno}},
  \bibinfo{journal}{Phys. Rev. Lett.} \textbf{\bibinfo{volume}{94}},
  \bibinfo{pages}{013202} (\bibinfo{year}{2005}).

\bibitem[{\citenamefont{Buchachenko et~al.}(2007)\citenamefont{Buchachenko,
  Cha{\l}asi{\'n}ski, and Szcz{\k{e}}{\'s}niak}}]{AABuchachenko:07}
\bibinfo{author}{\bibfnamefont{A.~A.} \bibnamefont{Buchachenko}},
  \bibinfo{author}{\bibfnamefont{G.}~\bibnamefont{Cha{\l}asi{\'n}ski}},
  \bibnamefont{and} \bibinfo{author}{\bibfnamefont{M.~M.}
  \bibnamefont{Szcz{\k{e}}{\'s}niak}}, \bibinfo{journal}{Eur. Phys. J. D}
  \textbf{\bibinfo{volume}{45}}, \bibinfo{pages}{147} (\bibinfo{year}{2007}).

\bibitem[{\citenamefont{Hutson and Green}(1994)}]{JMHutson:MOLSCAT14}
\bibinfo{author}{\bibfnamefont{J.~M.} \bibnamefont{Hutson}} \bibnamefont{and}
  \bibinfo{author}{\bibfnamefont{S.}~\bibnamefont{Green}},
  \emph{\bibinfo{title}{computer code {MOLSCAT}, version 14}},
  \bibinfo{howpublished}{CCP6, Daresbury} (\bibinfo{year}{1994}).

\bibitem[{\citenamefont{Gonz\'alez-Mart\'{i}nez and
  Hutson}(2007)}]{MLGonzalez-Martinez:07a}
\bibinfo{author}{\bibfnamefont{M.~L.} \bibnamefont{Gonz\'alez-Mart\'{i}nez}}
  \bibnamefont{and} \bibinfo{author}{\bibfnamefont{J.~M.}
  \bibnamefont{Hutson}}, \bibinfo{journal}{Phys. Rev. A}
  \textbf{\bibinfo{volume}{75}}, \bibinfo{pages}{022702}
  (\bibinfo{year}{2007}).

\bibitem[{\citenamefont{Hutson}(2011)}]{JMHutson:FIELD1}
\bibinfo{author}{\bibfnamefont{J.~M.} \bibnamefont{Hutson}},
  \emph{\bibinfo{title}{computer code {FIELD}, version 1}}
  (\bibinfo{year}{2011}).

\bibitem[{\citenamefont{{Le Roy} and Bernstein}(1970)}]{RJLeRoy:70a}
\bibinfo{author}{\bibfnamefont{R.~J.} \bibnamefont{{Le Roy}}} \bibnamefont{and}
  \bibinfo{author}{\bibfnamefont{R.~B.} \bibnamefont{Bernstein}},
  \bibinfo{journal}{J. Chem. Phys.} \textbf{\bibinfo{volume}{52}},
  \bibinfo{pages}{3869} (\bibinfo{year}{1970}).

\bibitem[{\citenamefont{Gribakin and Flambaum}(1993)}]{GFGribakin:93}
\bibinfo{author}{\bibfnamefont{G.~F.} \bibnamefont{Gribakin}} \bibnamefont{and}
  \bibinfo{author}{\bibfnamefont{V.~V.} \bibnamefont{Flambaum}},
  \bibinfo{journal}{Phys. Rev. A} \textbf{\bibinfo{volume}{48}},
  \bibinfo{pages}{546} (\bibinfo{year}{1993}).

\bibitem[{\citenamefont{Gao}(2004)}]{BGao:04}
\bibinfo{author}{\bibfnamefont{B.}~\bibnamefont{Gao}}, \bibinfo{journal}{J.
  Phys. B: At. Mol. Opt. Phys.} \textbf{\bibinfo{volume}{37}},
  \bibinfo{pages}{4273} (\bibinfo{year}{2004}).

\end{thebibliography}
\end{document}